\documentclass[sigconf]{acmart}
\usepackage{graphicx}
\usepackage{placeins}
\usepackage{comment}
\usepackage{float}
\usepackage{dblfloatfix}
\AtBeginDocument{%
  }

\setcopyright{acmlicensed}
\copyrightyear{2025}
\acmYear{2025}
\acmDOI{XXXXXXX.XXXXXXX}

\acmConference[SC'25]{The International Conference for High Performance Computing, Networking, Storage, and Analysis}{November 16--21, 2025}{St. Louis, MO}

\acmISBN{978-1-4503-XXXX-X/25/11}

\begin{document}

\title{Real-time ML-based Defense Against Malicious Payload in Reconfigurable Embedded Systems}

\author{Rye Stahle-Smith}
\affiliation{
  \institution{University of South Carolina}
  \city{Columbia}
  \state{South Carolina}
  \country{USA}
}
\email{ryes@email.sc.edu}

\author{Rasha Karakchi (Advisor)}
\affiliation{
  \institution{University of South Carolina}
  \city{Columbia}
  \state{South Carolina}
  \country{USA}
}
\email{karakchi@cec.sc.edu}

\renewcommand{\shortauthors}{Stahle-Smith and Karakchi}

\begin{abstract}
The growing use of FPGAs in reconfigurable systems introduces security risks through malicious bitstreams that could cause denial-of-service (DoS), data leakage, or covert attacks. We investigated chip-level hardware malicious payload in embedded systems and proposed a supervised machine learning method to detect malicious bitstreams via static byte-level features. Our approach diverges from existing methods by analyzing bitstreams directly at the binary level, enabling real-time detection without requiring access to source code or netlists. Bitstreams were sourced from state-of-the-art (SOTA) benchmarks and re-engineered to target the Xilinx PYNQ-Z1 FPGA Development Board. Our dataset included 122 samples of benign and malicious configurations. The data were vectorized using byte frequency analysis, compressed using TSVD, and balanced using SMOTE to address class imbalance. The evaluated classifiers demonstrated that Random Forest achieved a macro F1-score of 0.97, underscoring the viability of real-time Trojan detection on resource-constrained systems. The final model was serialized and successfully deployed via PYNQ to enable integrated bitstream analysis.
\end{abstract}

\keywords{PYNQ, Machine Learning (ML), AES-128, RS232, Hardware Trojan, Bitstreams, Truncated SVD, SMOTE}


\maketitle

\vspace{-1em}

\section{Introduction}
Known for delivering high-performance computation with lower power consumption than GPUs or multicore processors, FPGAs support dynamic updates, making them ideal for adaptive systems, real-time acceleration, and cloud environments. Major providers like Microsoft and IBM have integrated FPGAs into their cloud platforms for applications such as deep learning inference and accelerated networking \cite{elnaggar2023learning}. 

\begin{figure}[htbp]
  \centering
  \hspace*{-0.02\textwidth}
  \includegraphics[width=0.5\textwidth]{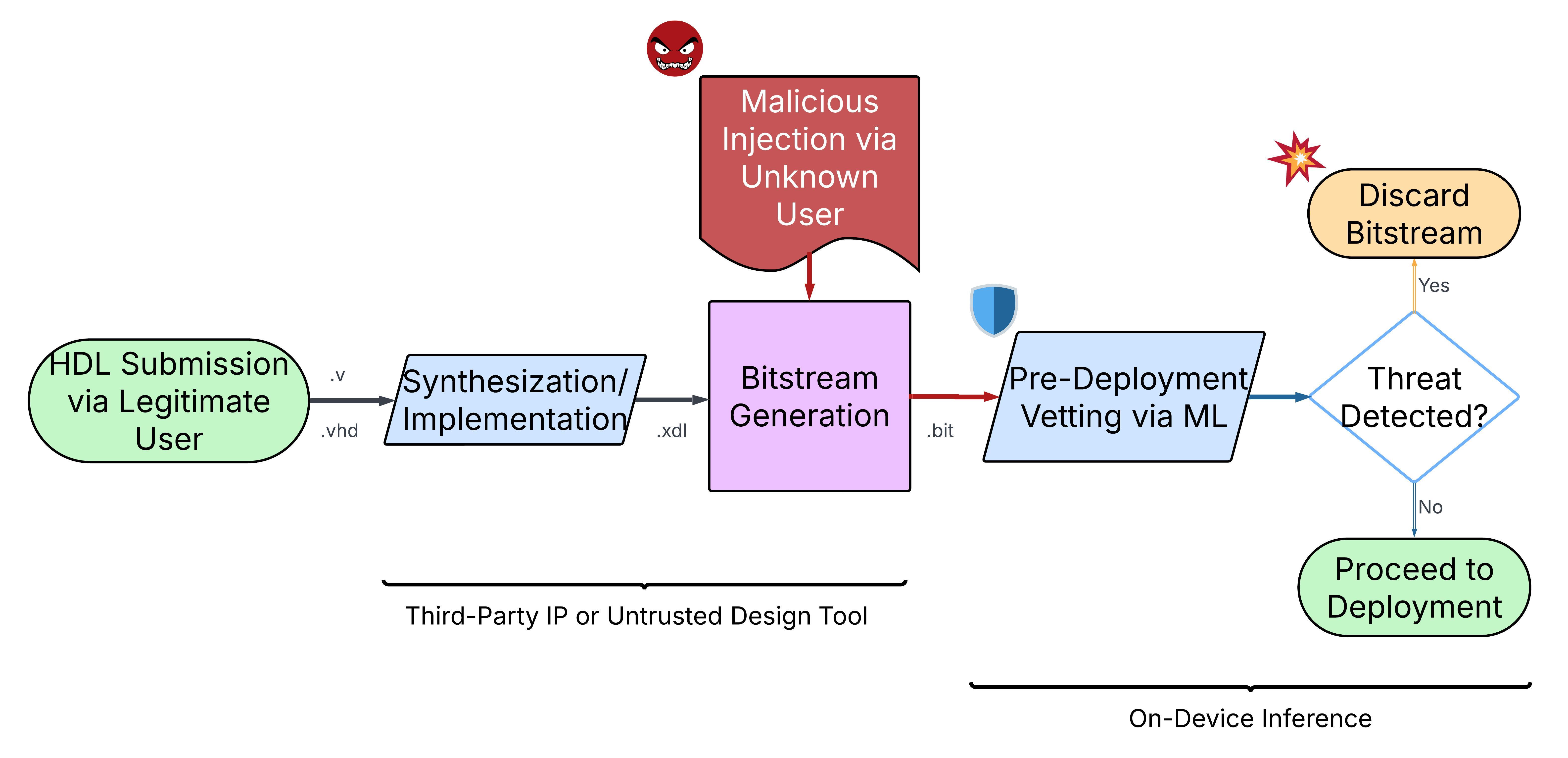}
  \caption{Threat Model for Detecting Trojan-Injected FPGA Bitstreams in Multi-Tenant Systems.}
  \label{fig:fig6}
\end{figure}
\vspace{-1em}

However, this flexibility also opens the door to new attacks. Since FPGAs are configured via binary bitstreams, malicious actors can reprogram devices with compromised data. This risk escalates in multitenant cloud setups, where hardware is shared through time-based reconfiguration, as illustrated in Figure \ref{fig:fig6}. Vendors traditionally mitigate risks by controlling synthesis pipelines or centrally generating bitstreams from submitted IP, but this raises privacy and usability issues, especially when IP confidentiality is critical. In contrast, prior work \cite{yoon2018bitstream, seo2018reverse, benz2012bil} focuses on detecting hardware Trojans by reverse-engineering bitstreams back into netlists for analysis, which can be time-consuming and reliant on availability of development-stage files.

Our approach diverges by analyzing bitstreams directly at the binary-level, enabling real-time detection without dependency on source or netlist availability. Open hardware initiatives offer transparency and community verification \cite{hayashi2025hardware}, yet hidden vulnerabilities persist, highlighting the need for binary-level detection methods. Prior work has applied machine learning to detect malicious FPGA configurations in shared, multi-tenant settings \cite{elnaggar2023learning}, leveraging structured feature extraction and layered classification to identify complex threats. This project builds on that foundation by targeting embedded environments, where computational overhead and memory are limited. Through streamlined byte-level analysis and efficient data representation, our approach enables real-time detection without reliance on attack-specific heuristics or multistage preprocessing. By analyzing bitstream structure, we aim to identify hardware Trojans predeployment. 

\vspace{-5pt}

\section{Methodology}
\begin{figure}[htbp]
  \centering
  \hspace*{0.01\textwidth}
  \includegraphics[width=0.5\textwidth]{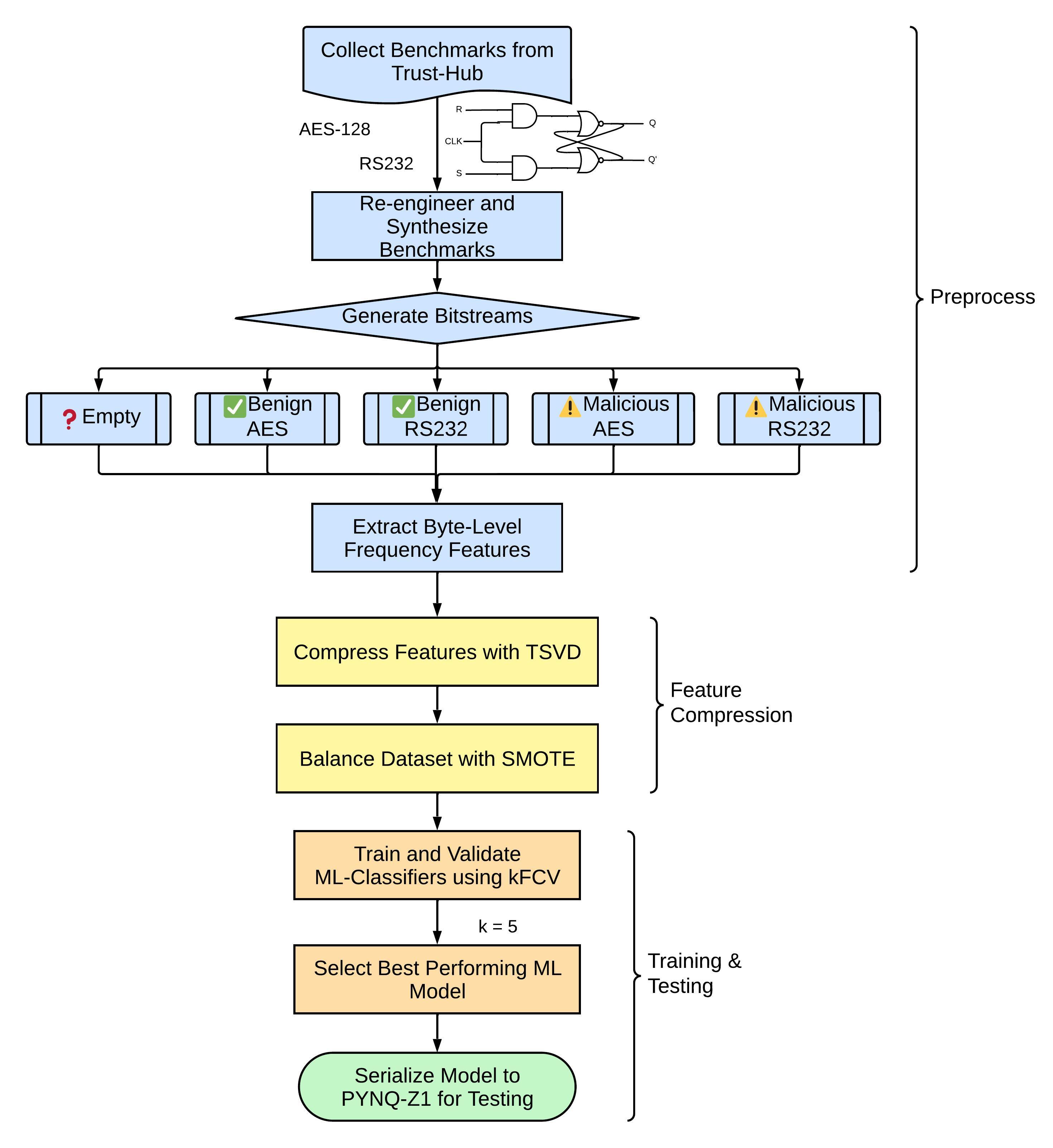}
  \caption{Training and Deployment Pipeline for Bitstream-Level Trojan Detection.}
  \label{fig:fig2}
\end{figure}

\begin{figure}[htbp]
  \centering
  \hspace*{0.05\textwidth}
  \includegraphics[width=0.3\textwidth]{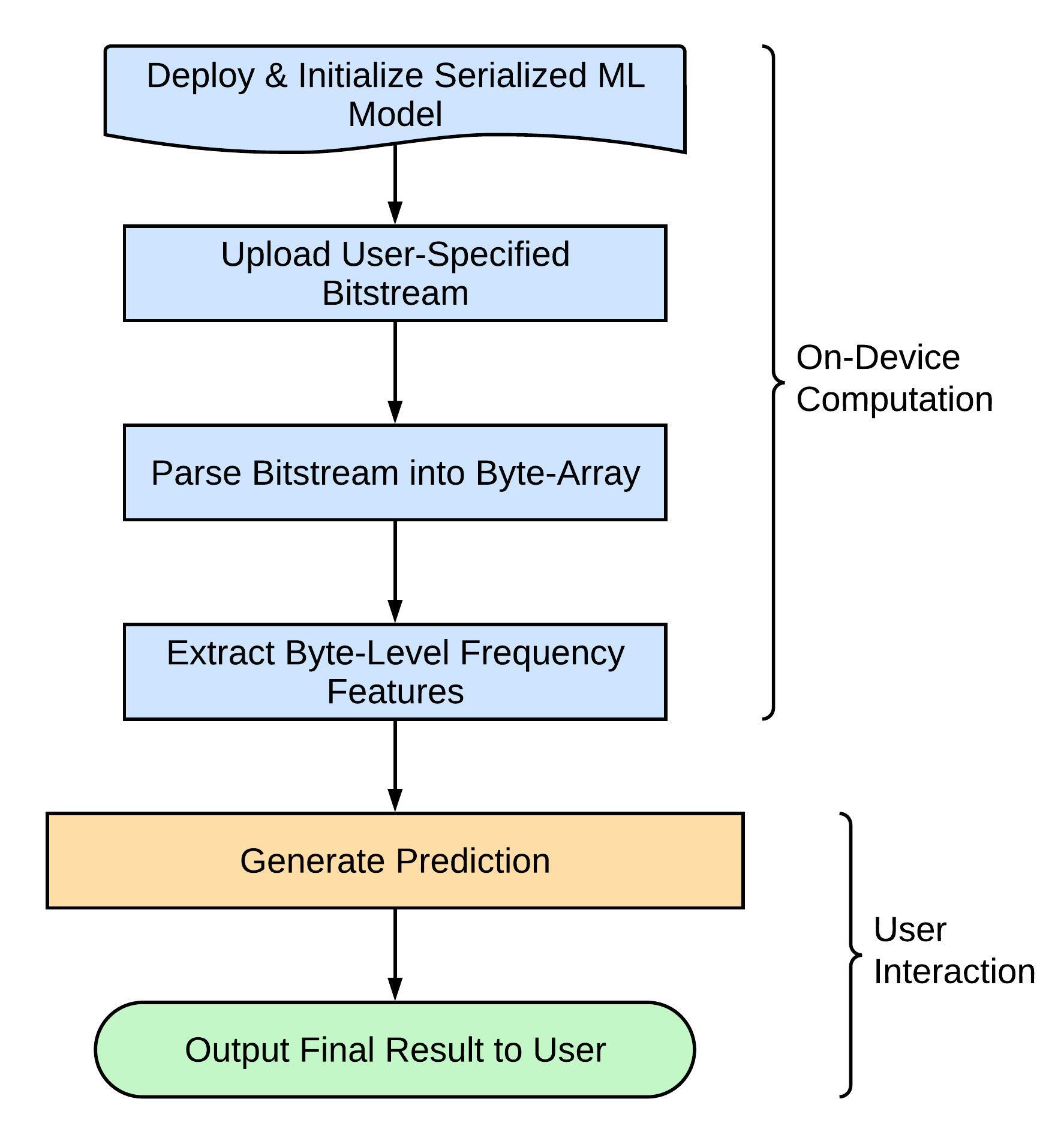}
  \caption{Runtime Prediction Pipeline on PYNQ-Z1 via Jupyter Notebook UI.}
  \label{fig:fig3}
\end{figure}

The general methodology is described in Figures \ref{fig:fig2} and \ref{fig:fig3}, which illustrate the complete lifecycle of our approach, from offline training and model selection to real-time deployment and inference. Our approach began with benchmark designs sourced from Trust-Hub \cite{shakya2017benchmarking, salmani2013design}, AES-128 and RS232 variants were synthesized into bit files and labeled as benign, malicious, or empty. Byte-level features were extracted from each file, followed by dimensionality reduction using Truncated Singular Value Decomposition (TSVD) to retain structural relevance while improving computational efficiency \cite{sklearn_tsvd}. 
To address data set imbalance and model bias \cite{elnaggar2018machine}, the Synthetic Minority Oversampling Technique (SMOTE) was applied to the minority classes in the sample before training \cite{chawla2002smote, imblearn_smote}. Multiple classifiers were evaluated with the detailed results summarized in Table \ref{tab:performance}.
These classifiers were assessed using k-Fold Cross-Validation (kFCV), a method that divides the dataset into k-subsets, systematically training and testing the model across different folds to robustly estimate performance \cite{sklearn_cross_validation}. This technique often includes mechanisms for parameter tuning and can leverage parallel computation to efficiently select the best-performing model \cite{pedregosa2011scikit}.

The best performing model was serialized and deployed to the PYNQ-Z1 \cite{pynq2024}, where it performed on-device inference using the same preprocessing pipeline to generate predictions. Table \ref{tab:classification-results} summarizes the inference performance and classification results across five deployment trials on the PYNQ-Z1, highlighting the model’s consistency and efficiency. This final step validated the feasibility of real-time Trojan detection in a constrained embedded environment without reliance on external computation.

\section{Results and Conclusion}
The proposed ML-based framework for FPGA bitstream classification achieved strong performance across multiple metrics. As shown in Table~\ref{tab:performance}, the Random Forest classifier reached an accuracy, precision, recall, and F1-score of $0.98 \pm 0.02$. On the hold-out test set, it achieved a true positive rate (TPR) of $97.14\%$, false negative rate (FNR) of $2.86\%$, false positive rate (FPR) of $0.8\%$, and an F1-score of $97\%$. 

\begin{table}[h!]
\footnotesize
\centering
\caption{Performance Metrics (Accuracy, Precision, Recall, F1 Score) for Evaluated Machine Learning Models.}
\label{tab:performance}
\begin{tabular}{|l|c|c|c|c|}
\hline
\textbf{Model} & \textbf{Accuracy} & \textbf{Precision} & \textbf{Recall} & \textbf{F1 Score} \\
\hline
Random Forest     & $0.98 \pm 0.02$ & $0.99 \pm 0.02$ & $0.98 \pm 0.02$ & $0.98 \pm 0.02$ \\
Gradient Boosting & $0.90 \pm 0.05$ & $0.88 \pm 0.11$ & $0.90 \pm 0.06$ & $0.88 \pm 0.09$ \\
AdaBoost          & $0.89 \pm 0.05$ & $0.87 \pm 0.10$ & $0.89 \pm 0.05$ & $0.87 \pm 0.08$ \\
Logistic Regression & $0.33 \pm 0.07$ & $0.14 \pm 0.03$ & $0.36 \pm 0.08$ & $0.20 \pm 0.05$ \\
Naive Bayes       & $0.95 \pm 0.06$ & $0.96 \pm 0.05$ & $0.96 \pm 0.05$ & $0.96 \pm 0.05$ \\
SVM (RBF)         & $0.46 \pm 0.10$ & $0.30 \pm 0.13$ & $0.51 \pm 0.11$ & $0.36 \pm 0.12$ \\
KNN               & $0.86 \pm 0.04$ & $0.87 \pm 0.04$ & $0.86 \pm 0.03$ & $0.86 \pm 0.04$ \\
Decision Tree     & $0.95 \pm 0.00$ & $0.96 \pm 0.00$ & $0.95 \pm 0.00$ & $0.95 \pm 0.00$ \\
\hline
\end{tabular}
\end{table}

The confusion matrix confirmed perfect classification across five classes. The model was successfully deployed on the PYNQ-Z1 platform, with Table~\ref{tab:classification-results} showing consistent results over five trials and an average prediction latency of $3.35$ seconds. These findings validate the effectiveness of the approach for real-time Trojan detection in resource-constrained environments. Future work will aim to improve robustness against evasive threats and reduce false positives without sacrificing accuracy.

\begin{table}[htbp]
\footnotesize
\caption{Classification and Timing per Trial}
\label{tab:classification-results}
\begin{tabular}{@{}c c c r r r@{}}
\toprule
\textbf{Trial} & \textbf{Actual} & \textbf{Predicted} & \textbf{Load} & \textbf{Extract} & \textbf{Predict} \\
 & \textbf{Class} & \textbf{Class} & \textbf{(ms)} & \textbf{(ms)} & \textbf{(ms)} \\
\midrule
1 & Mal. RS232 & Mal. RS232 & 239.48 & 3157.72 & 15.03 \\
2 & Mal. RS232 & Mal. RS232 & 190.72 & 3268.73 & 16.94 \\
3 & Ben. AES   & Ben. AES   &  23.65 & 3167.01 & 16.81 \\
4 & Ben. AES   & Ben. AES   &  21.85 & 3156.55 & 16.96 \\
5 & Mal. AES   & Mal. AES   & 189.02 & 3228.77 & 15.85 \\
\bottomrule
\end{tabular}
\end{table}









\section{Acknowledgments}
  This work was supported by the McNair Junior Fellowship and OUR at the University of South Carolina. This project used benchmark designs from Trust-Hub, a resource sponsored by the National Science Foundation (NSF). OpenAI’s ChatGPT was employed to help refine language and grammar. All technical content and analysis were developed independently by the authors. This research also made use of the PYNQ-Z1 FPGA platform, provided by AMD and Xilinx, whose tools and hardware enabled the synthesis and deployment stages of this study.

\bibliographystyle{apalike}
\bibliography{references.bib}

\end{document}